\documentclass[aps,pre,preprint,groupedaddress]{revtex4-1}
\usepackage{graphics,graphicx}
\usepackage{amsmath}
\usepackage{epstopdf}

\newcommand{\bb}[1]{\left[ #1 \right]}
\newcommand{\bp}[1]{\left( #1 \right)}

\newcommand{\ve}[1]{\vert #1 \vert}

\begin{document}


\title{Extended Order Parameter and Conjugate Field for the Dynamic Phase Transition in a Ginzburg-Landau Mean-Field Model in an Oscillating Field}


\author{Daniel T. Robb}
\affiliation{Department of Mathematics, Computer Science and Physics, Roanoke College, Salem, VA 24153}
\affiliation{Department of Physics, Astronomy and Geology, Berry College, Mount Berry, GA 30149}
\author{Aaron Ostrander}
\affiliation{University of Maryland, College Park, MD 20742, USA}
\affiliation{Department of Physics, Astronomy and Geology, Berry College, Mount Berry, GA 30149}


\date{\today}

\begin{abstract}
We present numerical evidence for an extended order parameter and conjugate field for the dynamic phase transition in a Ginzburg-Landau mean-field model driven by an oscillating field. The order parameter, previously taken to be the time-averaged magnetization, comprises the deviations of the Fourier components of the magnetization from their values at the critical period. The conjugate field, previously taken to be the time-averaged magnetic field, comprises the even Fourier components of the field. The scaling exponents $\beta$ and $\delta$ associated with the extended order parameter and conjugate field are shown numerically to be consistent with their values in the equilibrium mean-field model.

\end{abstract}


\maketitle

\section{Introduction}

Dynamic phase transitions (DPTs) have been identified in a variety of physical systems, and can serve as valuable aids in understanding non-equilibrium systems. A well-studied DPT in magnetic systems occurs when the period of an applied oscillating magnetic field of sufficient amplitude drops below a critical period $P_c$, causing the symmetric hysteresis loop to bifurcate into two asymmetric loops \cite{Tome90, Mendes91, Acharyya99}. Below the critical period, the DPT in magnetic systems has been shown in mean-field models \cite{Fujisaka01} and kinetic Ising model simulations \cite{Sides98,Sides99,Korniss00} to exhibit critical scaling with the same critical exponent $\beta$ as the corresponding equilibrium transitions, with the period-averaged magnetization serving as a dynamic order parameter \cite{Fujisaka01,Sides98,Sides99}. Recent work has shed light on the behavior in the critical region \cite{Idigoras12,Gallardo12}, examined the dependence on the stochastic dynamics \cite{Buendia08}, and investigated the DPT in novel theoretical \cite{Baez10, Deviren10, Park13, Kinoshita10} and experimental \cite{Deviren12} contexts.

In numerical simulations of the two-dimensional Ising model in an oscillating field, it was shown that the period-averaged magnetic field serves as a field conjugate to the dynamic order parameter in the two-dimensional Ising model \cite{Robb07}. Evidence for a DPT in an Ising-like experimental magnetic system, using the period-averaged magnetic field as the dynamic order parameter, was provided in Ref. \cite{Robb08}. However, this recent work did not establish that the period-averaged magnetic field (called $H_b$ in Refs.~\cite{Robb07} and \cite{Robb08}) is the \textit{only} component of the conjugate field. For example, the same results would have been found in Ref.~\cite{Robb07} if the full conjugate field $H_c$ were actually $H_c = H_b + H_d$, where $H_d$ is a function of the applied field which happened to be zero in all cases studied in Refs.~\cite{Robb07} and \cite{Robb08}. 

Here we study a particular \textit{mean-field model} and demonstrate numerically that, at least in the case of the mean-field model chosen, there are indeed additional components to the conjugate field. We also demonstrate that there are additional components to the dynamic order parameter, at least near the critical period $P_c$. We speculate that similar results will hold for the kinetic Ising model and other driven, spatially-extended models, but do not provide evidence for such models in this paper. 

\section{Computational Model}

The mean field model studied here has the Ginzburg-Landau (GL) free energy $F(m) = am^2 + bm^4 - h m$, where the magnetization $m=m(t)$ and magnetic field $h=h(t)$ are time-dependent but spatially uniform. This produces the dynamical equation
\begin{equation}
\frac{\mathrm{d}m}{\mathrm{d}t} = -\frac{\partial F}{\partial m} = -2am - 4bm^3 + h ,
\label{eqn-motion}
\end{equation}
which is a more general form of Eq.~(3) governing the spatially uniform solutions in Ref.~\cite{Fujisaka01}. It is known and straightforward to show that the equilibrium critical exponents for this mean-field Ginzburg-Landau (MFGL) model are $\beta = 1/2$ and $\delta = 3$. The dynamic critical exponents for this MFGL model at the critical period match the corresponding exponents for the  equilibrium transition, as they do for the kinetic Ising model studied in Ref.~\cite{Robb07}. We believe this result for the MFGL model has been demonstrated previously; at least the dynamical exponent $\beta = 1/2$ is established in Eq.~(23) of Ref~\cite{Fujisaka01}. In any case, we establish the dynamic critical exponents $\beta = 1/2$ and $\delta = 3$ numerically in Figs.~\ref{DPT_epsilon_scaling} and \ref{h_scaling}, respectively, of this paper.

In implementing the MFGL, we choose parameters $a=-\frac{3\sqrt{3}}{4}$ and $b=\frac{3\sqrt{3}}{8}$, which for $h=0$  yield minima of the free energy at $m = \pm 1$. In a periodic applied field $h(t) = h(t+P)$, we expect the dynamics to converge to limit cycle(s) of the form $m(t) = m(t+P)$. Setting $\omega = 2\pi/P$, we can expand both $h(t)$ and $m(t)$ as complex Fourier series:
\begin{equation}
h(t) = \sum_k h_k e^{ik\omega t} \; \; ; \; \; m(t) = \sum_k m_k e^{ik\omega t}
\label{mh-expansion}
\end{equation}
where here and for the remainder of this paper, a summation index without limits is understood to run from $-\infty$ to $+\infty$. Since $h(t)$ and $m(t)$ are real, it follows that $h_{-k} = h_k^*$ and $m_{-k} = m_k^*$, so that $h_0$ and $m_0$ are real. The dynamic order parameter referred to as $Q$ in previous studies \cite{Sides98,Sides99,Korniss00} is the real Fourier coefficient $m_0$, while the component of the conjugate field identified in Ref. \cite{Robb07} -- the period-averaged magnetic field -- is the real Fourier coefficient $h_0$.  

\section{Higher-order bifurcations}

\begin{figure}[t]
\includegraphics[width=4.0in]{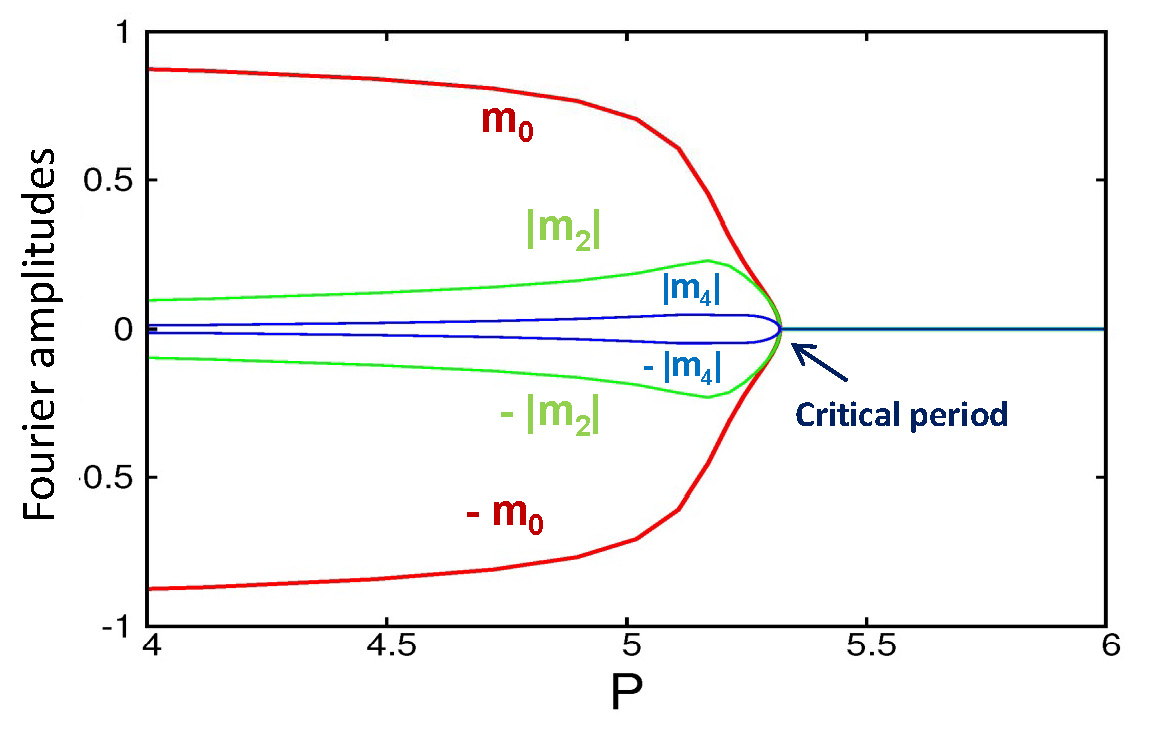}
\caption{\label{Dynamic_phase_diagram} \textbf{(Color online)} Dynamic phase diagram illustrating the bifurcation of Fourier coefficients $m_0$, $m_2$ and $m_4$ below $P_C$. Here $a=-\frac{3\sqrt{3}}{4}$, $b=\frac{3\sqrt{3}}{8}$, and $H_1 = 1.5$, for which it is found that $P_C = 5.319357661995$. Note that the values $+\ve{m_{2j}}$ and $-\ve{m_{2j}}$ are displayed below $P_c$ in Fig.~\ref{Dynamic_phase_diagram} for simplicity; the two stable asymmetric loops actually have opposite complex Fourier components $m_{2j}$ and $-m_{2j}$. }
\end{figure}

In both mean-field \cite{Tome90,Fujisaka01} and kinetic Ising \cite{Sides98,Sides99,Korniss00} models, above $P_c$ there is a stable symmetric hysteresis loop with $m_0 = 0$. Below $P_c$ there are two stable asymmetric loops with opposite values $m_0 = \pm m_s$, as well as one unstable symmetric loop with $m_0=0$. This behavior of $m_0$ in the GL model defined by Eq.~(\ref{eqn-motion}) can be observed in Fig.~\ref{Dynamic_phase_diagram}. In addition, the Fourier components $m_2$ and $m_4$ undergo a similar bifurcation at $P_c$. That is, above $P_c$ there is a stable symmetric hysteresis loop with $m_2 = m_4 = 0$. Below $P_c$ there are two stable asymmetric loops with opposite values $m_2 = \pm m_{s,2}$ and $m_4 = \pm m_{s,4}$, as well as one unstable symmetric loop with $m_0=0$.  It was shown in Ref. \cite{Fujisaka01} that $m_{2j} = 0$ (for $j$ integer) above $P_c$, but the bifurcation of $m_2$ and $m_4$ below $P_c$ has not been reported before to our knowledge. A similar bifurcation occurs for all even Fourier components $m_{2j}$. It is interesting to note, however, that whereas the constant component $m_0$ increases monotonically below $P_c$, the amplitudes $|m_2|$ and $|m_4|$ increase over a limited range below $P_c$, and then decrease, eventually approaching 0 as the period $P$ decreases to 0.

To within the numerical accuracy of our simulations, the bifurcation in all three components $m_0, m_2$ and $m_4$ occurred at the same critical value $P_c$. The critical period can be located numerically by applying the stability criterion
\begin{equation}
\sum_k \ve{m_{k,c}}^2 = - \frac{a}{6b}
\label{critical.period.criterion}
\end{equation}
along the line of solutions with $m_0 = 0$. Here the notation $m_{k,c}$ refers to the Fourier components of the steady state magnetization $m(t)$ at the critical period $P_c$. To establish Eq.~(\ref{critical.period.criterion}), we follow Ref.~\cite{Fujisaka01} and perturb Eq.~(\ref{eqn-motion}) around the stable solution $m(t)$, giving to first order  $\frac{\mathrm{d}}{\mathrm{d}t}\bb{\delta m(t)} = -2a \bb{\delta m(t)} - 12b\bb{m(t)}^2 \bb{\delta m(t)}$. This has solution $\delta m(t) = \delta m(0) \exp \bb{-\int_0^t \bp{2a+12b\bb{m(t')}^2} \mathrm{d}t'}$. Evaluating at $t=P$, we find that the perturbation will grow, i.e., the solution $m(t)$ is unstable, if $-\int_0^P \bb{2a+12b\bb{m(t')}^2}\mathrm{d}t' > 0$. Expanding the two factors of $m(t')$ in their Fourier components using Eq.~(\ref{mh-expansion}), this can be shown to be equivalent to the condition $\sum_{k=-\infty}^{k=+\infty} \ve{m_{k}}^2 < - \frac{a}{6b}$, which establishes Eq.~(\ref{critical.period.criterion}). For the parameters used here ($a=-\frac{3\sqrt{3}}{4}$, $b=\frac{3\sqrt{3}}{8}$), and with a sinusoidal applied field $h(t) = H_1 \cos(\omega t)$ with $H_1 = 1.5$, the critical period was determined using Eq.~(\ref{critical.period.criterion}) to be $P_c = 5.319357661995$.

The bifurcation in the even Fourier components $m_{2j}$ can be understood using Fourier analysis. We assume that the driving field $h(t)$ contains (arbitrary) odd Fourier components $h_k$, including a non-zero $h_1$. Inserting the expansions in Eq.~(\ref{mh-expansion}) into Eq.~(\ref{eqn-motion}) yields (for all integer $k$)
\begin{equation}
0 = -(i\omega  k + 2 a) m_k - 4b\displaystyle\sum_{n_1,n_2} m_{n_1}m_{n_2}m_{k-n_1-n_2} + h_k 
\label{eom-fourier}
\end{equation}
For odd $k$, the terms in the sum in Eq.~(\ref{eom-fourier}) must contain either 0 or 2 even Fourier components $m_k$. (Here `even Fourier component' refers to a Fourier component with even index.) Thus, the equations for odd $k$ are still satisfied if the signs of all the even Fourier components $m_k$ are reversed. For even $k$, each term in the sum in Eq.~(\ref{eom-fourier}) must contain either 1
or 3 even Fourier components $m_k$. By the above assumption, $h_k$ = 0 in this case, so changing the sign of all $m_k$ will reverse the sign of all terms in the equation, and the equations for even $k$ also remain satisfied. Thus, stable loops below $P_c$ come in pairs. The two stable loops in the pair share the same value for the odd $m_k$, and values with opposite signs for the even $m_k$ values.

\section{Scaling with respect to period}

We investigated numerically the scaling of both odd and even Fourier components $m_k$ below the critical period $P_c$. Because we investigate deviations in various quantities at and nearby a \textit{numerically determined} critical period, this requires very accurate simulation, achieved using Cash-Karp Runge-Kutta integration in long double precision variables (accurate to twenty decimal places on the computer system used). The steady-state loops for a given field period $P$ above $P_c$ were determined using a shooting method, which located the initial magnetization values $m(0)$ resulting in the same subsequent value $m(t=P) = m(0)$ at the end of the field cycle. The use of the shooting method circumvents the issue of critical slowing down occurring near the critical period, in which the convergence time to the steady-state becomes inconveniently large during normal time evolution. For the scaling variables, we use the scaled period $\epsilon = \frac{P_c - P}{P_c}$ and 
\begin{equation}
z_k = \sqrt{\ve{m_k}^2 - \ve{m_{k,c}}^2}
\label{z_k_definition}
\end{equation}
Note the scaling variable $z_k$ reduces to $\ve{m_k}$ for even $k$, since $m_{k,c} = 0$ in this case.

\begin{figure}[t]
\includegraphics[width=4.0in,angle=0]{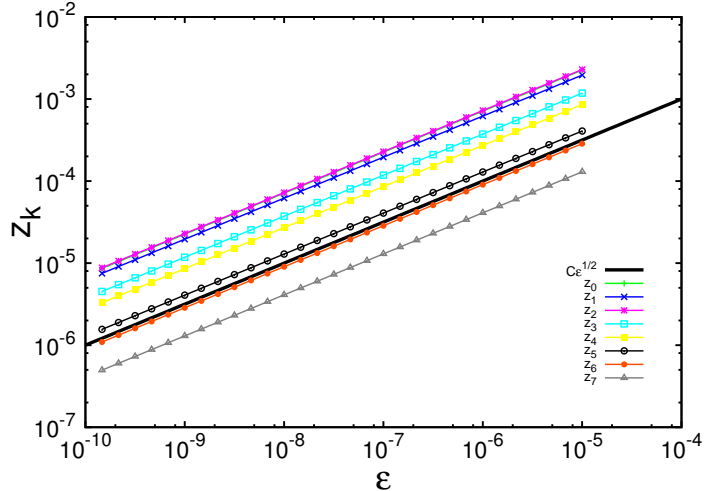}
\caption{\label{DPT_epsilon_scaling} \textbf{(Color online)} Critical scaling of order parameter scaling variables $z_k$ with respect to the scaled period $\epsilon$. The plots for $z_k$, $k=0,...7$ are shown individually in the figure, along with a reference line representing scaling with exponent 1/2.} 
\end{figure}

As shown in Fig.~\ref{DPT_epsilon_scaling}, the quantities $z_k$ scale with respect to the scaled period $\epsilon$ with critical exponent 1/2, for Fourier components $k=0$ through $k=7$. This agrees with the scaling exponent ($\beta = 1/2$) previously determined for $m_0$ in the GL model (see Eq.~(23) of Ref. \cite{Fujisaka01}). We have explicitly confirmed the scaling with exponent $\beta = 1/2$ up to index $k=40$. (We are confident the scaling continues with exponent $\beta = 1/2$ beyond $k=40$. However, since $z_k$ decreases with $k$, as seen in Fig.~\ref{DPT_epsilon_scaling}, the values of $z_k$ decrease below the accuracy of our numerical simulation past $k=40$.) 

Defining the deviation $\delta m_k = m_k - m_{k,c}$, it is straightforward to show that the fact that $z_k \sim \epsilon^{1/2}$ implies that $\delta m_k \sim \epsilon$ for $k$ odd, and $\delta m_k \sim \epsilon^{1/2}$ for $k$ even. This scaling of the deviations can be confirmed analytically using a perturbation of the Fourier relation in Eq.~(\ref{eom-fourier}) for frequency $\omega = \omega_c + \delta \omega$, i.e., just below the critical period. We insert $m_k = m_{k,c} + \delta m_k$ into Eq.~(\ref{eom-fourier}), expand and group terms, and then subtract Eq.~(\ref{eom-fourier}) with the critical values $m_{k,c}$. Noting to first order $\delta \omega = \epsilon \omega_c$, the result is
\begin{widetext}
\begin{eqnarray}
0 = - i \epsilon \omega_c k m_{k,c} - (i\omega_c k + 2a) \delta m_k - 12b\displaystyle\sum_{n_1,n_2} m_{n_1,c} m_{n_2,c} \delta m_{k-n_1-n_2} & \nonumber \\*
- 12b\displaystyle\sum_{n_1,n_2} m_{k-n_1-n_2,c} \delta m_{n_1} \delta m_{n_2} &  - 4b\displaystyle\sum_{n_1,n_2} \delta m_{n_1} \delta m_{n_2} \delta m_{k-n_1-n_2}
\label{eom-fourier2}
\end{eqnarray}
\end{widetext}

If we assume scaling relationships of the form
\begin{equation}
  \delta m_k =\begin{cases}
    c_k \epsilon^{p}, & \text{for k odd}\\
    c_k \epsilon^{q}, & \text{for k even}
  \end{cases}
\label{scaling-form}
\end{equation}
then the scaling exponents $p=1$ and $q=1/2$ can be determined from Eq.~(\ref{eom-fourier2}) as follows. Considering Eq.~(\ref{eom-fourier2}) for odd $k$, for example $k=1$, the first term $-i \epsilon \omega_c m_{1,c}$ is linear in $\epsilon$. The rest of the terms (to lowest order in the deviations for even and odd $k$) must then be linear in $\epsilon$ as well, in order that the equation obtained by inserting the scaling forms in Eq~(\ref{scaling-form}) is independent of $\epsilon$. The first sum $\displaystyle\sum_{n_1,n_2} m_{n_1,c} m_{n_2,c} \delta m_{1-n_1-n_2}$ involves only odd $\delta m_k$, since $n_1$ and $n_2$ must be odd in order that the term in the sum be nonzero. This establishes that the scaling exponent $p = 1$ for the odd terms. The second sum $\displaystyle\sum_{n_1,n_2} m_{1-n_1-n_2,c} \delta m_{n_1} \delta m_{n_2}$ has non-zero terms with $n_1$ and $n_2$ either both odd or both even. If $n_1$ and $n_2$ are both odd, the term scales as $\epsilon^2$ and can be neglected. The terms with both $n_1$ and $n_2$ even are the lowest order terms including the even $\delta m_k$, and must scale linearly in $\epsilon$, implying that the scaling exponent $q=1/2$. The third sum is higher order in both even and odd $\delta m_k$ and can be neglected for critical scaling. It is also interesting to note that the system of equations represented by Eq.~(\ref{eom-fourier2}) has a solution with all even $\delta m_k = 0$. In this case, the set of equations for odd $k$ forms (to lowest order) a linear system whose solution is the \textit{unstable} symmetric loop below $P_c$.

\section{Scaling with respect to field components}

\begin{figure}[h]
\includegraphics[width=4.0in]{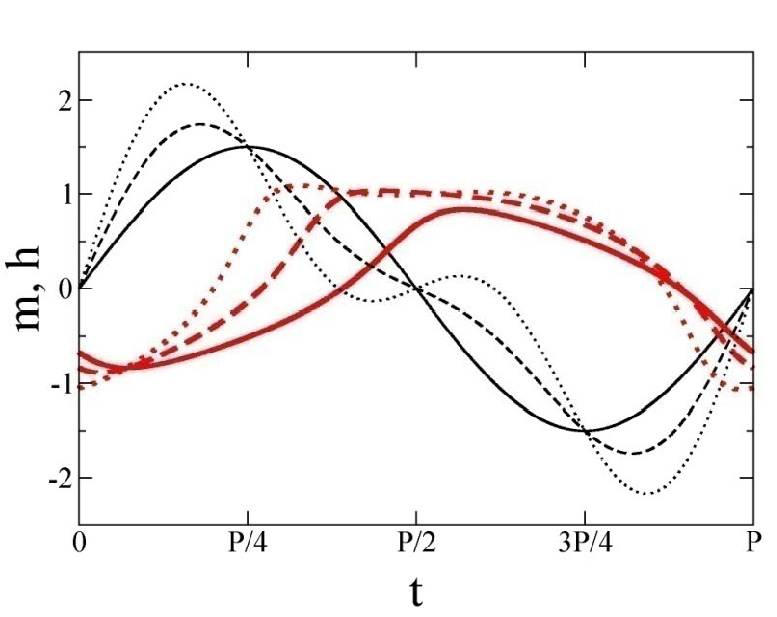}
\caption{\label{Effect_of_h2} \textbf{(Color online)}  Plot of $h(t)$ and  $m(t)$ for GL model with applied field $h(t) = H_1 \sin\bp{\omega t} + H_2 \sin\bp{2\omega t}$, with $H_1 = 1.5$ and $\omega = \frac{2\pi}{P_c}$. The thin curves (black online) represent $h(t)$; the thick curves (red online) represent $m(t)$. The cases $H_2 = 0.0, 0.5$ and 1.0 are represented by solid, dashed and dotted curves, respectively.}
\end{figure}

We next provide numerical evidence that all $h_j$ (for $j$ even) are components of the conjugate field, which yield the same scaling exponent associated with $h_0$. First, though, it is helpful to consider a specific case in which even Fourier components other than $h_0$ can produce a non-zero value of $m_0$, as this may seem counterintuitive. In Fig.~\ref{Effect_of_h2}, the magnetization and field are plotted as a function of time for the applied field $h(t) = H_1 \sin\bp{\omega t} + H_2 \sin\bp{2\omega t}$, with $H_1 = 1.5$ and increasing values of the amplitude $H_2$. With $H_2 = 0$, we find $h(t+P/2)=-h(t)$ and $m(t+P/2)=-m(t)$, respectively, which implies $h_0 = 0$ and $m_0 = 0$. With $H_2 = 0.5$, the maximum of the field, and therefore the maximum of the magnetization, occurs earlier in the cycle. Due to the hysteresis in the model, the system spends a greater percentage of the field cycle with positive magnetization, producing a value $m_0 > 0$. With $H_2 = 1.0$, the maxima of the field and the magnetization occur even earlier in the field cycle, producing an even larger value of $m_0$.

\begin{figure}[h]
\includegraphics[width=4.0in,angle=0]{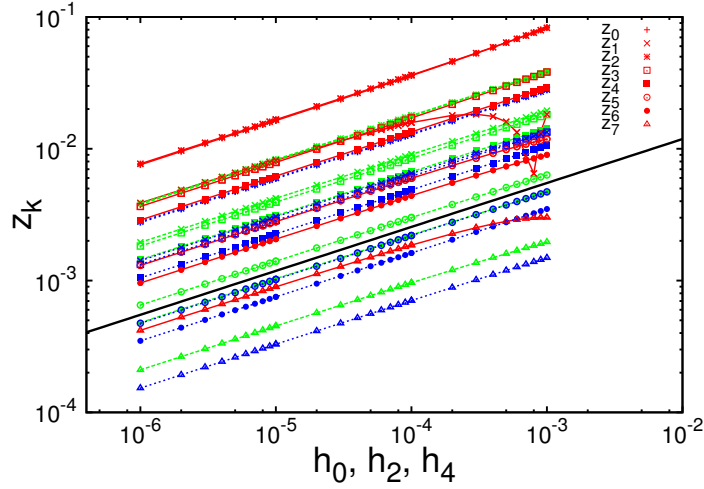}
\caption{\label{h_scaling} \textbf{(Color online)} Critical scaling of the variables $z_k$ ($k=0,..7$) with respect to $h_0$, $h_2$ and $h_4$. The scaling with respect to $h_0$, $h_2$ and $h_4$ are represented by solid lines (red online), dashed lines (green online), and dotted lines (blue online), respectively. The black reference line shows exact scaling with exponent 1/3.}
\end{figure}

\begin{figure}[h]
\includegraphics[width=4.0in,angle=0]{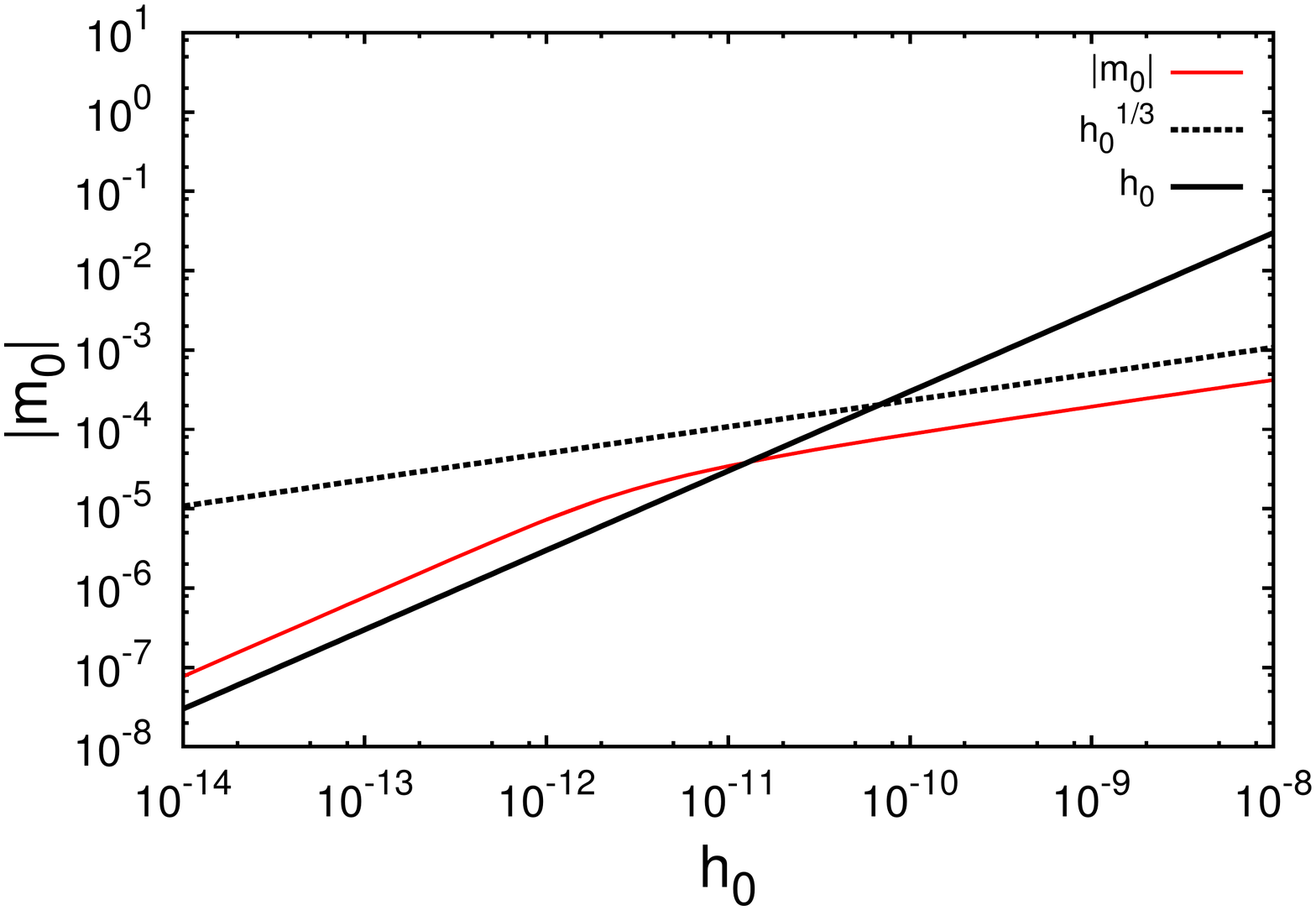}
\caption{\label{h0_scaling_crossover} \textbf{(Color online)} Crossover of critical scaling of $z_0=|m_0|$ with respect to $h_0$, at the period value $P=5.3193577 > P_c$. The data for $z_0$ is represented by the thin line (red online). The thick black dashed curve represents scaling with exponent 1/3, while the thick black dash-dotted curve represents scaling with exponent 1.}
\end{figure}

We next investigate whether, as suggested by Fig.~\ref{Effect_of_h2}, other even Fourier components of the magnetic field function as parts of the conjugate field. For the scaling variables associated with the conjugate field, we use the Fourier components $h_j$ ($j$ even). For the scaling variables associated with the order parameter, we again use the quantities $z_k$ defined in Eq.~(\ref{z_k_definition}). In Fig.~\ref{h_scaling}, we observe that all of the variables $z_k$ ($k=0,..7$) exhibit critical scaling with exponent 1/3 with respect to the amplitudes $h_0$, $\ve{h_2}$ and $\ve{h_4}$ of the zeroth, second and fourth Fourier coefficients of the applied field. In each case, we have explicitly confirmed the scaling with exponent 1/3 up to index $k=50$ (the limit of our numerical accuracy). In addition, the scaling of $z_k$ ($k=0,..7$) with respect to $h_j$ ($j$ even) with exponent 1/3 has been explicitly verified up to $j=30$ (the limit of our numerical accuracy).  The critical exponent agrees with that found for mean-field models for the scaling of the magnetization with respect to the field at the critical temperature ($1/\delta = 1/3$). We emphasize that Fig.~\ref{h_scaling} illustrates the interesting fact that \textit{each} scaling variable $z_0,z_1,z_2...$ (and its associated magnetization component $m_0,m_1,m_2$ exhibit scaling independently with respect to \textit{each} even field component $h_0,h_2,h_4...$. We have not examined the effect of changes to more than one field component simultaneously.

\begin{figure}[h]
\includegraphics[width=4.0in,angle=0]{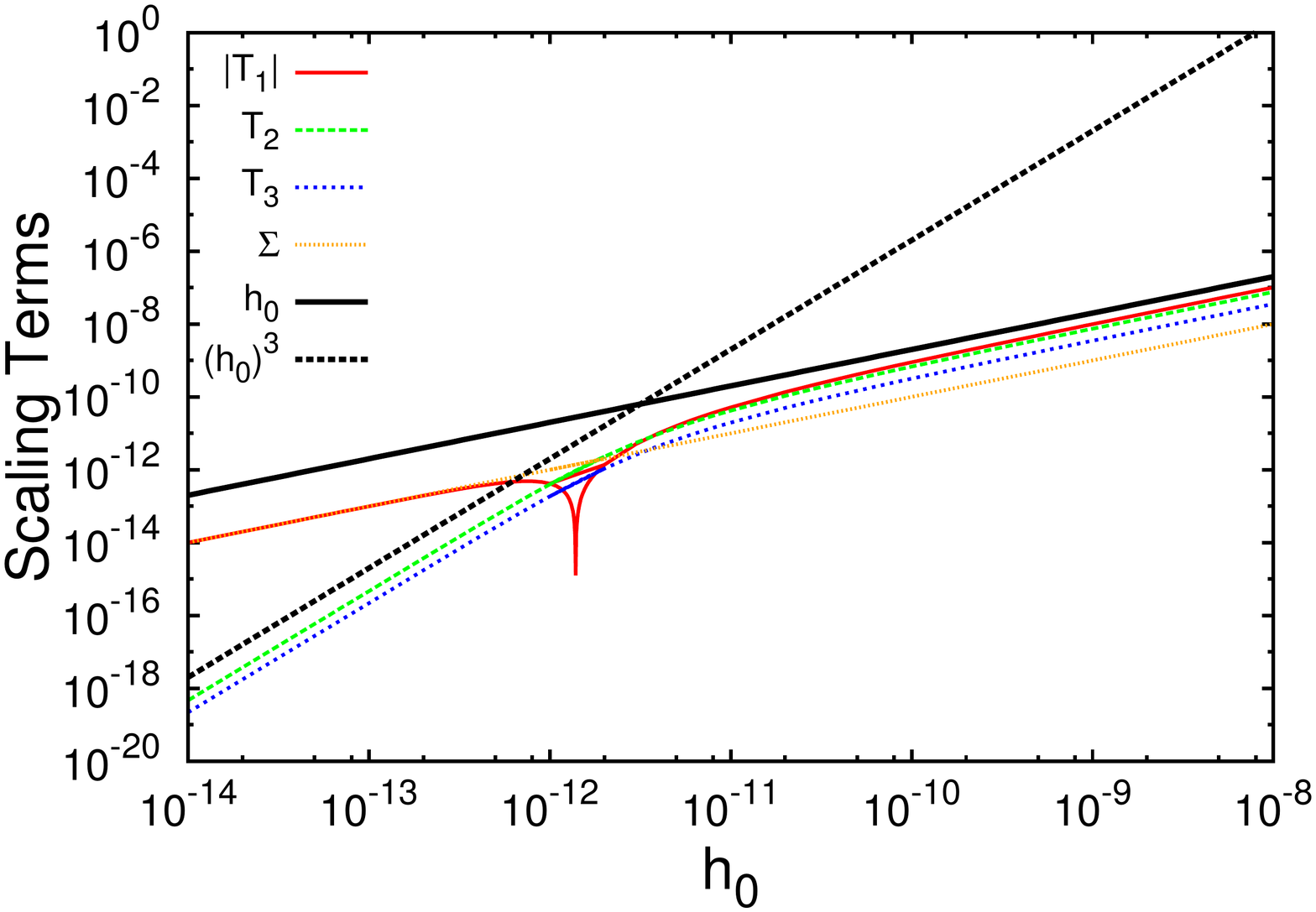}
\caption{\label{h0_scaling_terms} \textbf{(Color online)} Illustration of the three bracketed terms ($T_1$, $T_2$, $T_3$) from Eq.~(\ref{eom-fourier3}) with respect to $h_0$, at the period $P=5.3193577$. The signs of $T_2$ and $T_3$ remain positive throughout. The sign of $T_1$ switches from positive to negative just above $h_0 = 10^{-12}$. Due to the logarithmic scale used, the absolute value $|T_1|$ is therefore plotted. The sum of the three terms, denoted $\Sigma = T_1 + T_2 + T_3$, is also shown. The thick black dashed curve shows represents scaling with exponent 1, while the thick black dotted curve represents scaling with exponent 3.}
\end{figure}

Note that a change in an odd Fourier component of the applied field ($\delta h_j$, $j$ odd) serves only to relocate the critical period, with the relative shift $\epsilon = \frac{{P_{C}'-P_C}}{P_C} \sim \delta h_j$ (the direction of the shift changing with the sign of $\delta h_j$). As a result, introducing a change $\delta h_j$ ($j$ odd) at $P_c$ will (through the shift in the critical period) bring about a change $z_k \sim \epsilon^{1/2} \sim |\delta h_j|^{1/2}$ for $k$ odd. If the critical period is decreased by the change $\delta h_j$, then $z_k = 0$ for $k$ even will be zero. However, if the critical period is increased by $\delta h_j$, $z_k$ for $k$ even will also scale as $z_k \sim \epsilon^{1/2} \sim |\delta h_j|^{1/2}$.

We can understand several important aspects of these numerical scaling results with respect to the field by considering the analogue of Eq.~(\ref{eom-fourier2}) for the case in which perturbations $\delta h_k$ to the field's Fourier components are introduced:\\

\begin{widetext}
\begin{eqnarray}
0 = \left[ 2a \delta m_k - 12b\displaystyle\sum_{n_1,n_2} m_{n_1,c} m_{n_2,c} \delta m_{k-n_1-n_2} \right] & \nonumber \\*
+ \left[ - 12b\displaystyle\sum_{n_1,n_2} m_{k-n_1-n_2,c} \delta m_{n_1} \delta m_{n_2} \right] & + \left[- 4b\displaystyle\sum_{n_1,n_2} \delta m_{n_1} \delta m_{n_2} \delta m_{k-n_1-n_2} \right] + \delta h_k
\label{eom-fourier3}
\end{eqnarray}
\end{widetext}

As an example, consider a perturbation with $\delta h_0 = h_0$, and the other $\delta h_k = 0$ (for $k \neq 0$). We examine the scaling behavior of $z_0 = |m_0|$ with respect to $h_0$ at a period $P = 5.3193577$, just above the critical period $P_c = 5.319357661995$. As seen in Fig.~(\ref{h0_scaling_crossover}), the scaling of $|m_0|$ undergoes a crossover from linear scaling ($\sim h_0$) to cube root scaling ($\sim h_0^{1/3}$) in the range from $h_0 = 10^{-12}$ to $h_0 = 10^{-11}$. Simulations at values of $P$ closer to $P_c$ show that the crossover region moves to progressively lower values of $h_0$ as $P$ approaches $P_c$, so that the scaling at $P_c$ has exponent 1/3, as previously illustrated in Fig.~\ref{h_scaling}.\\

Fig.~\ref{h0_scaling_terms} illustrates the behavior of the three bracketed terms ($T_1,T_2,T_3$)  in Eq.~(\ref{eom-fourier3}), as well as their sum. The interaction of the three terms in creating the crossover from linear to cube root scaling in Fig.~\ref{h0_scaling_crossover} is somewhat involved, but can be understood in general terms as follows. Again taking $k=0$ as a specific example, when $\delta h_0 = h_0$ is very small, the resulting deviations $\delta m_k$ will be very small. Thus, the term $T_1$ linear in $\delta m_k$ dominates, while the much smaller $T_2$ and $T_3$ scale with a higher power ($\sim (h_0)^{3}$).  As $h_0$ grows, the values $\delta m_k$ increase, and the sums in terms $T_2$ and $T_3$ become comparable in size to $T_1$. In addition, the sum within $T_1$ finally dominates the single term $2a \delta m_k$, so that $T_1$ crosses from positive to negative. Past this point, all three terms $T_1$, $T_2$, and $T_3$ scale linearly with $h_0$, as seen in Fig.~\ref{h0_scaling_terms}. Given that the term $T_3$, comprised of three-term products of the deviations $\delta m_k$, scales linearly with $h_0$, the relationship $\delta m_k \sim (h_0)^{1/3}$, seen for the case $k=0$ in Fig.~\ref{h0_scaling_crossover}, is then determined. In addition, note that as $P$ approaches $P_C$, the coefficient of $\delta m_k$ within the term $T_1$, i.e.,
$$2a - 12b \sum_{n_1+n_2=0} m_{n_1,c} m_{n_2,c} = 2a - 12b \sum_{k} \vert m_{k,c}\vert^2 $$
approaches zero, as may be seen from the condition for $P_c$ in Eq.~(\ref{critical.period.criterion}). Thus, the switch of $T_1$ from positive to negative, and the accompanying crossover from linear to cube root scaling, occurs at smaller and smaller values of $h_0$ as $P$ approaches $P_c$. \\

\section{Conclusion and Future Work}

We have verified that analogous scaling results are seen starting with a square-wave field or a triangular wave field, which each consist of a particular set of odd Fourier components $h_j$, rather than the sinusoidal field (only $h_1$) used here. That is, \textit{each} scaling variable $z_0,z_1,z_2...$, consisting of deviations from the values associated with the basic applied field form, exhibits scaling independently with respect to \textit{each} even field component $h_0,h_2,h_4...$ added to the basic applied field form. Given this, it is reasonable to hypothesize that the set of odd Fourier components $h_j$ determine a dynamic phase transition with critical period and unstable symmetric loops below the critical period; the even Fourier components of the field then serve as components of a conjugate field in this dynamic phase transition. 

It would be interesting to determine if a single composite conjugate field can be constructed from the even Fourier components $h_j$, at least near the critical period, which would require investigating the effect of introducing several even Fourier components of the field simultaneously. Given that higher order magnetization components $m_2, m_4,...$ do not increase monotonically below $P_c$ (as seen in Fig.~\ref{DPT_epsilon_scaling}), such a single composite conjugate field would likely be limited to the immediate neighborhood of the critical period $P_c$. 

Finally, while the MFGL model we have used does capture the basic physics of the ferromagnetic phase transition, spatially dependent models such as the kinetic Ising model, as well as more specific models of particular geometries (e.g., superlattices, multilayers or nanostructures), are of more practical interest. We speculate that similar extensions of order parameter and conjugate field will occur in some form in these more realistic systems, but it is important and worthwhile to test this directly, and to discover what practical importance these higher-order components of the dynamic order parameter and conjugate field may have.

\begin{acknowledgments}
We would like to acknowledge Mark Novotny and Per Arne Rikvold for introducing us to the fascinating phenomenon of the dynamic phase transition, and would like to thank Andreas Berger for several useful comments on the present manuscript.
\end{acknowledgments}


\providecommand{\noopsort}[1]{}\providecommand{\singleletter}[1]{#1}%

\end{document}